\documentclass[aps,preprint]{revtex4}

\usepackage{amssymb}
\usepackage{amsmath}
\usepackage{epsfig}
\usepackage{epstopdf}
\usepackage{bm}
\usepackage{graphicx,epsfig}
\usepackage{mathrsfs}
\usepackage{dcolumn}
\usepackage{color}
\usepackage{natbib}
\usepackage{CJK}
\def\be{\begin{equation}}
\def\ee{\end{equation}}
\def\bea{\begin{eqnarray}}
\def\eea{\end{eqnarray}}

\allowdisplaybreaks[2]

\begin{document}

\title{Synthetic space with arbitrary dimensions in a few rings undergoing dynamic modulation}
\author{Luqi Yuan$^{1}$, Meng Xiao$^{1}$, Qian Lin$^2$ and Shanhui Fan$^1$}
\affiliation{$^1$Department of Electrical Engineering, and Ginzton
Laboratory, Stanford University, Stanford, CA 94305, USA \\
$^2$Department of Applied Physics, Stanford University, Stanford,
CA 94305, USA}

\date{\today }

\begin{abstract}
We show that a single ring resonator undergoing dynamic modulation
can be used to create a synthetic space with an arbitrary
dimension. In such a system the phases of the modulation can be
used to create a photonic gauge potential in high dimensions. As
an illustration of the implication of this concept, we show that
the Haldane model, which exhibits non-trivial topology in two
dimensions, can be implemented in the synthetic space using three
rings. Our results point to a route towards exploring
higher-dimensional topological physics in low-dimensional physical
structures. The dynamics of photons in such synthetic spaces also
provides a mechanism to control the spectrum of light.
\end{abstract}


\maketitle

\section{Introduction}

It is well known that the property of a physical system depends
critically on its dimension. Moreover, it has been recognized that
for a lattice system, its dimension can be controlled by designing
the coupling constants, i.e. the connectivity, between different
sites. This has led to the development of the concept of synthetic
dimension, where one explores the physics in higher dimensions
using lower-dimensional physical structures
\cite{tsomokos10,boada12,jukic13,boada15,suszalski16,taddia17,celi14,luo14,qi08,kraus13,lian16,lian17}.

As one significant recent development, it has been recognized that
ring resonators provide a natural platform to explore various
synthetic dimension concepts
\cite{yuanOL,yuanOptica,ozawa16,qianweyl}. A ring resonator
supports multiple resonant modes at different frequencies. By
modulating the ring dynamically, there modes can couple with one
another, forming a lattice system along the frequency axis. With a
particular choice of the modulation format, a ring resonator,
which can be thought of conceptually as a zero-dimensional object,
is described by a one-dimensional tight binding model.
Consequently, one can use a $N$-dimensional array of dynamically
modulated rings to study the physics in $N+1$ dimension.

In this paper, we take a step further in the development of the
concept of synthetic dimension using ring resonators. In a
dynamically modulated ring resonator, the coupling between
different modes is controlled by the modulation format
\cite{yuanOptica}. We show that a choice of modulation format that
is different from Ref. \cite{yuanOL,yuanOptica,ozawa16,qianweyl}
can enable us to use a single ring to explore physics at
\textit{arbitrary} dimensions. And moreover, the phase degrees of
freedom in the modulation naturally lead to a photonic gauge
potential and associated non-trivial topology in such higher
dimensional synthetic dimension \cite{yuanOL}. As an example, we
show that one can implement the Haldane model \cite{haldane88},
which is an important model with non-trivial topology in
two-dimensions, using only three dynamically modulated resonators.
Our work points to a route of exploring the very rich fundamental
physics of higher dimensional system in practical physical
structures. The use of the frequency axis in this implementation
also leads to enhanced capability for manipulating the spectrum of
light.

The paper is organized as follows: In Sec. II, we show that with a
properly chosen modulation format, a single ring undergoing
dynamic modulation can be mapped into a Hamiltonian that describes
a lattice in arbitrary dimensions. In Sec. III, building upon the
concept developed in Sec. II, we show that we can realize the
Haldane model using only three ring resonators undergoing dynamic
modulation. In Sec. IV, we provide a detailed simulations of the
proposal in Sec. III, and highlight the experimental signature
associated with this proposal. We conclude our paper in Sec. V.

\section{Single ring resonator supporting a synthetic two-dimensional space}

\begin{figure}[h]
\centering
\includegraphics[width=0.8\linewidth]{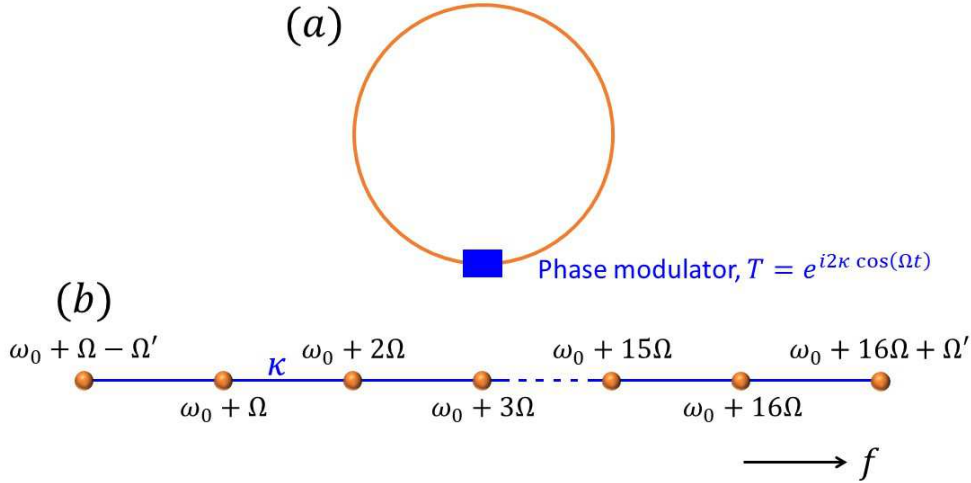}
\caption{(a) A single ring resonator undergoes dynamic modulation.
(b) The structure in (a) can be described by a one-dimensional
tight-binding model with nearest-neighbor coupling along the
frequency dimension. \label{Fig:scheme}}
\end{figure}

We start with a single ring resonator consisting of a single-mode
waveguide. Such a ring resonator supports multiple resonant modes.
For the $m$-th resonant mode, its wavevector along the waveguide
$\beta_m$ needs to satisfy:
\begin{equation}
\left( \beta_m - \beta_0  \right) L = 2\pi m, \label{Eq1:res}
\end{equation}
where $L$ is the circumference for the ring. Assuming zero group
velocity dispersion in the waveguide, the resonant frequency is
\begin{equation}
\omega_{m} = \omega_0 + m \Omega, \label{Eq1:resfreq}
\end{equation}
where $\Omega$ is the free-spectral-range of the ring. Suppose we
place in the ring a phase modulator, which has a time-dependent
transmission coefficient
\begin{equation}
T= e^{i 2 \kappa \cos (\Omega t + \phi)}, \label{Eq4:T1}
\end{equation}
where we choose the modulation frequency to be equal to  $\Omega$
and the modulation strength $\kappa$. $\phi$ is the modulation
phase. In the presence of an incident wave $e^{i\omega t}$, the
transmitted wave has the form $Te^{i\omega t} = e^{i\omega t+i 2
\kappa \cos (\Omega t + \phi)}$, which to the lowest order  of
$\kappa$ can be approximated as $Te^{i\omega t} \approx e^{i\omega
t} + i \kappa \left[e^{i(\omega +\Omega)t + i \phi} + e^{i(\omega
-\Omega)t -  i \phi}\right]$. Therefore, this system can be
described by the Hamiltonian
\begin{equation}
H = \sum_m \omega_m a^\dagger_m a_m +  \sum_m  \left[2\kappa
\cos(\Omega t + \phi) a^\dagger_m a_{m+1} + h.c.\right],
\label{Eq1:H}
\end{equation}
where $a_m(a^\dagger_m)$ is the annihilation (creation) operator
for the $m$-th mode. With $c_m \equiv a_m e^{-i\omega_m t}$, Eq.
(\ref{Eq1:H}) becomes
\begin{equation}
H_{RWA} =  \sum_m \left(\kappa c^\dagger_m c_{m+1} e^{i\phi} +
h.c.\right), \label{Eq1:Hrwa}
\end{equation}
under the rotating wave approximation. The ring resonator under
modulation in the form of Eq. (\ref{Eq4:T1})  is therefore
described by the simple tight-binding model with nearest-neighbor
coupling along the frequency dimension \cite{yuanOL,yuanOptica}
(Figure \ref{Fig:scheme}(b)). As have been noted in Refs.
\cite{yuanOL,yuanOptica,ozawa16}, the modulation phase $\phi$ is a
gauge potential for photon.  For the discussion in this section we
will set $\phi$ to zero for simplicity.

When the group velocity dispersion is not zero, the modes in the
ring are no longer equally spaced and hence the modulation no
longer induces on-resonance coupling. The group velocity
dispersion thus results in a natural ``boundary'' in the frequency
space \cite{yuanOL}. As an illustration, in Figure
\ref{Fig:scheme}(b), by choosing $\omega_1 - \omega_0 =
\omega_{17}-\omega_{16} = \Omega' \neq \Omega$, we form a finite
one-dimensional lattice with 16 resonant modes.

\begin{figure}[h]
\centering
\includegraphics[width=\linewidth]{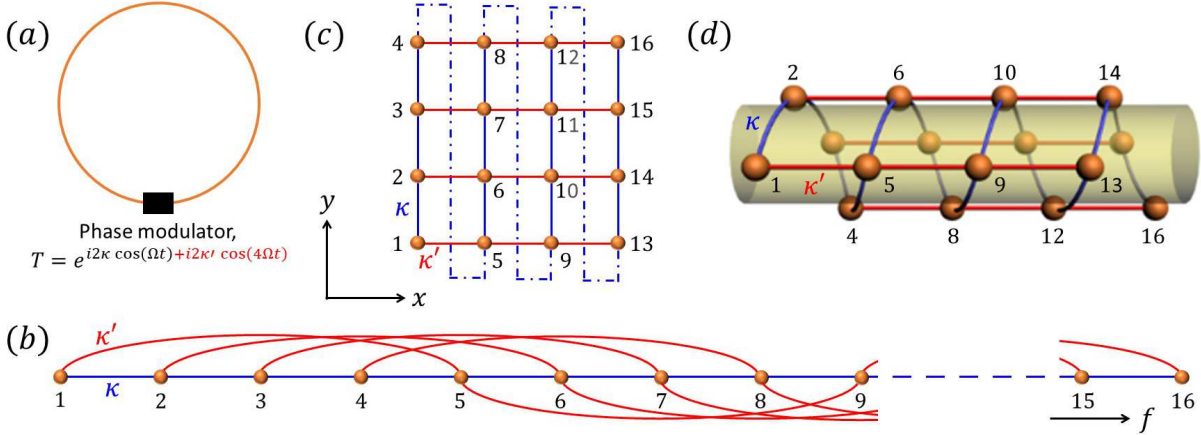}
\caption{(a) A single ring resonator undergoes dynamic modulation
with a format as described by Eq. (\ref{Eq4:T2}) with $N = 4$. (b)
Such a design supports a one-dimensional model along the frequency
dimension with the long-range connectivity. (c) Such a
one-dimensional model is equivalent to a two-dimensional $4\times
4$ square lattice structure with additional connections labelled
by dashed-dotted curves. (d) The lattice in (c) can alternatively
be represented by a two-dimensional lattice on a cylindrical
surface with twisted boundary condition. in (c) and (d), the blue
and red lines correspond the coupling induced by the first and
second terms in Eq. (\ref{Eq4:T2}), respectively.
\label{Fig:ring2D}}
\end{figure}

We now show that a more complex lattice geometry in the frequency
space can be achieved by choosing other modulation formats. As an
illustration, consider instead a phase modulator with the
following transmission coefficient:
\begin{equation}
T= e^{i \left[ 2 \kappa \cos (\Omega t) + 2 \kappa' \cos (N\Omega
t) \right]}, \label{Eq4:T2}
\end{equation}
where $N$ is an integer. The additional modulation term with a
modulation frequency $N\Omega$ provides a long-range coupling
along the frequency axis between the $m$-th and $m+N$-th modes
with the strength $\kappa'$ (see Figure \ref{Fig:ring2D}(b)). Thus
the system is now described by the Hamiltonian
\begin{equation}
H =  \sum_m \left( \kappa c^\dagger_m c_{m+1} + \kappa'
c^\dagger_m c_{m+N} + h.c.\right). \label{Eq1:H1D}
\end{equation}

The Hamiltonian in Eq. (\ref{Eq1:H1D}) in fact represents a
two-dimensional system with a twisted boundary condition. To
illustrate this in details, we provide an alternative graphic
representation of this Hamiltonian in Figure \ref{Fig:ring2D}(c),
for the case of $N = 4$, in the ring with the underlying modal
structure shown in Figure \ref{Fig:scheme}(b) which has a boundary
along the frequency axis. In this representation, the long-range
coupling as described by the second term in Eq. (\ref{Eq1:H1D}),
is represented by bonds along the $x$-direction. Most of the
short-range coupling as described by the first term in Eq.
(\ref{Eq1:H1D}) is represented by bonds along the $y$-direction,
with the exception of the coupling between the $4n$ and $4n+1$
modes, which are represented by a ``twisted'' coupling between the
lower and upper edge of the lattices. This ``twisted'' coupling
can also be represented by Figure \ref{Fig:ring2D}(d) where we see
that a ring modulated with two commensurate modulation frequency
can be represented by a two-dimensional lattice arranged on the
surface of  cylinder with a twisted periodic boundary condition.
Such a geometry has a screw symmetry along the axis of the
cylinder.

As illustrated above, a single ring resonator under the dynamic
modulation with two modulation frequencies corresponds to a
two-dimensional synthetic space. One can synthesize a
higher-dimensional lattice by adding additional modulation
frequencies.

\section{Realization of the Haldane model in the two-dimensional synthetic space with ring resonators }

In the previous section, we have shown that by choosing a
modulation format having two commensurate modulation frequencies,
one can use a single ring resonator to achieve a synthetic
two-dimensional lattice. In previous works
\cite{yuanOL,yuanOptica,ozawa16,qianweyl}, it was also noted that
the phase of the modulation corresponds to a gauge potential in
the synthetic space. In this section, as the main illustration of
the paper, we show that one can combine these two ideas to realize
the Haldane model using only three resonators.

\begin{figure}[!t]
\centering
\includegraphics[width=0.69\linewidth]{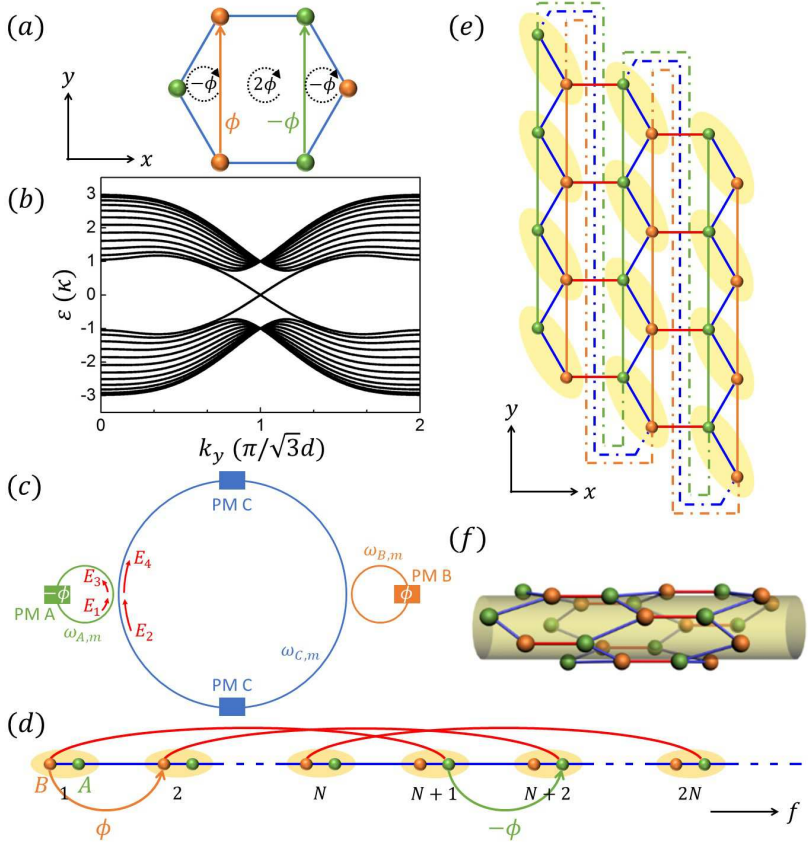}
\caption{(a) A single unit cell of the Haldane model consists of a
honeycomb lattice with nearest neighbor coupling, and the coupling
between the next nearest neighbors along the $y$-directions. which
carries a direction dependent coupling phase. (b) The projected
band structure for the Haldane model of (a) in a stripe geometry.
The stripe is infinite along the $y$-direction and has 12 unit
cells along the $x$ direction. (c) Two ring resonators $A$ and $B$
with resonant frequencies $\omega_{A,m}$ and $\omega_{B,m}$
respectively. Each ring has a phase modulator (PM) inside it.
In-between, there is an auxiliary ring $C$ with resonant
frequencies $\omega_{C,m}$ with two phase modulators inside it.
(d) A one-dimensional synthetic lattice along the frequency axis
is created by the dynamic modulators placed in rings as shown in
(c). Each unit cell of the lattice consists of a pair of the
$m$-th resonant mode $\left(\omega_{A,m},\omega_{B,m}\right)$ in
the $A$ and $B$ rings. (e) The one-dimensional model in (d) is
equivalent to a two-dimensional honeycomb lattice structure with
additional connections labelled by dashed-dotted curves. (f) The
two-dimensional lattice in (e) can be represented by a honeycomb
lattice on a cylindrical surface with a twisted boundary
condition. \label{Fig:haldane}}
\end{figure}

The Haldane model is one of the most important models that
exhibits non-trivial topological effects \cite{haldane88}. There
have been several pervious works implementing the Haldane model in
photonics \cite{jotzu14,xiao15,minkov16}, all of which relies upon
complex lattice structures. To show that the physics of the
Haldane model can be implemented in synthetic space, here we seek
to implement a Hamiltonian on the honeycomb lattice:
\begin{equation*}
H_h = \sum_{\vec r} \left( \kappa c^\dagger_{B,\vec r} c_{A,\vec r
+ \vec e_1} + \kappa c^\dagger_{B,\vec r} c_{A,\vec r + \vec e_2}
+ \kappa c^\dagger_{B,\vec r} c_{A,\vec r + \vec e_3} + h.c.
\right)
\end{equation*}
\begin{equation}
+  \sum_{\vec r} \left( \kappa' e^{-i\phi} c^\dagger_{A,\vec r}
c_{A,\vec r + \vec e_1 -\vec e_2} + \kappa' e^{i\phi}
c^\dagger_{B,\vec r} c_{B,\vec r + \vec e_1 -\vec e_2} + h.c.
\right)  ,\label{Eq4:Hhaldane}
\end{equation}
where $\vec e_1 = (-d/2, \sqrt{3}d/2)$, $\vec e_2 = (-d/2,
-\sqrt{3}d/2)$, $\vec e_3 = (d, 0)$, and $d$ is the lattice
constant. A single unit cell for this Hamiltonian is shown in
Figure \ref{Fig:haldane}(a). The subscripts $A$ and $B$ denote the
$A$ and $B$ sites of the honeycomb lattice, respectively. The
terms in the first parenthesis in Eq. (\ref{Eq4:Hhaldane})
describes nearest-neighbor coupling on a honeycomb lattice with a
coupling strength $\kappa$. The terms in the second parenthesis in
Eq. (\ref{Eq4:Hhaldane}) describe coupling between two
next-nearest-neighbor $A(B)$ sites along the $y$-direction, with a
coupling strength $\kappa'$ and a phase $-\phi(+\phi)$.

\newpage

This model is a slight variant of the original model in Ref.
\cite{haldane88}. The key feature of the Haldane model, which is
the presence of a local magnetic field that averages to zero in
each unit cell, is preserved in this Hamiltonian. As a result, the
two  models have very similar topologically non-trivial band
structures. To illustrate the topological properties of Eq.
(\ref{Eq4:Hhaldane}), we consider a stripe that is  infinite along
the $y$ axis (which makes $k_y$ a good quantum number) and has
$12$ unit cells along the $x$ axis. We set $\kappa'=\kappa/2$ and
$\phi=\pi/2$. In the bandstructure plotted in Figure
\ref{Fig:haldane}(b), one can see that there is a topologically
nontrivial gap between the upper (and lower) bulk bands, which has
Chern number +1 and -1, respectively, as evidenced by a pair of
topologically-protected one-way edge states inside the gap. Such a
band structure is very similar to that of the original Haldane
model in Ref. \cite{haldane88}. Therefore, for the rest of the
paper, we refer to the Hamiltonian of Eq. (\ref{Eq4:Hhaldane}) as
the Haldane model.

To implement the Hamiltonian in Eq. (\ref{Eq4:Hhaldane}), we
consider the geometry as shown in Figure \ref{Fig:haldane}(c). The
geometry consists of two rings having the same circumference $L$,
labelled as $A$ and $B$ respectively. Each ring has a phase
modulator inside it. The two rings also couple with each other
through an auxiliary ring $C$ which has two phase modulators in
it. Ring $A$ has modes with resonant frequencies $\omega_{A,m} =
\omega_0 + m \Omega$. These modes corresponds to the $A$ sites in
the Haldane model. Ring $B$ has modes with resonant frequencies
$\omega_{B,m} = \omega_0 - \Omega/4 + m \Omega$, corresponding to
the $B$ sites. Such a mode structure can be achieved, for example,
by choosing an appropriate waveguide design for the ring $B$, such
that its dispersion relation is a constant shift in wavevector
space with respect to that of the waveguide forming ring $A$, i.e.
$\beta_B (\omega) = \beta_A (\omega+\Omega/4)$. Each pair of the
$m$-th resonant mode $\left(\omega_{A,m},\omega_{B,m}\right)$ in
two rings comprises a unit cell (labelled as a pair of $A$ and $B$
sites) along the one-dimensional frequency axis in Figure
\ref{Fig:haldane}(d).

The modulator in ring $A(B)$ has a transmission coefficient
\begin{equation}
T_{A(B)}= e^{i  2 \kappa' \cos (\Omega t \pm \phi) },
\label{Eq4:TAB}
\end{equation}
which couples two nearest-neighbor resonant modes of type $A(B)$.
In-between two rings, we place an auxiliary ring $C$ that couples
rings A and B together, as shown in Figure \ref{Fig:haldane}(c).
The circumference of the auxiliary ring $C$ is set to be $4L$,
which gives the resonance condition $\left( \beta_m - \beta_0
\right) 4L = 2\pi m$. The dispersion relation for the waveguide
forming the ring $C$ is chosen as $\beta_C (\omega) = \beta_A
(\omega+\Omega/8)$. Hence the resonant frequency for the ring $C$
is $\omega_{C,m} = \omega_0 - \Omega/8 + m \Omega / 4$. Therefore,
the modes in both rings $A$ and $B$ are not resonant with the
modes in ring $C$. Ring $C$ is coupled with ring $A$ with the
coupling matrix
\begin{equation}
\left( {\begin{array}{*{20}c}
  E_3 \\
  E_4 \\
\end{array}} \right) =  \left( {\begin{array}{*{20}c}
  \sqrt{1-\gamma^2} & -i\gamma \\
  -i\gamma & \sqrt{1-\gamma^2} \\
\end{array}} \right) \left( {\begin{array}{*{20}c}
  E_1 \\
  E_2 \\
\end{array}} \right),
\label{Eq7:Tcouple}
\end{equation}
where $\gamma$ is the coupling strength and $E_i$ ($i=1,2,3,4$) is
the electric field amplitudes in both rings labelled in Figure
\ref{Fig:haldane}(c). Ring $C$ is also coupled with ring $B$ with
the same coupling matrix. On each arm of ring $C$, we place a
modulator with a time-dependent transmission coefficient:
\begin{equation}
T=  e^{i \left[ 2\kappa_1 \cos(\Omega t/4) + 2\kappa_1
\cos(3\Omega t/4) + 2\kappa_1 \cos(N\Omega t + \Omega t/4)
\right]}. \label{Eq4:Tbetween}
\end{equation}
Each modulator is localized at a distance $L$ away from the
couplers between the rings. Referring to the one-dimensional
lattice in Figure \ref{Fig:haldane}(d), the modulation with
frequency $\Omega /4$ provides the intra-cell coupling between
modes inside each unit cell, the modulation with frequency
$3\Omega /4$ provides the inter-cell coupling between $A$ and $B$
types of modes in neighbor cells, and the modulation with
frequency $N\Omega + \Omega /4$ provides a long-range coupling
between the $B$ site in the $m$-th unit cell and the $A$ site in
the $(m+N)$-th unit cell. This system can therefore be described
by the Hamiltonian
\begin{equation*}
H = \sum_m \left( \kappa c^\dagger_{B,m} c_{A,m} + \kappa
c^\dagger_{A,m} c_{B,m+1} + \kappa c^\dagger_{B,m} c_{A,m+N}
\right.
\end{equation*}
\begin{equation}
\left. + \kappa' e^{-i\phi} c^\dagger_{A,m} c_{A,m+1} + \kappa'
e^{i\phi} c^\dagger_{A,m} c_{B,m+1} + h.c. \right) ,
\label{Eq4:H1Dhaldane}
\end{equation}
where $\kappa= \kappa_1 \gamma^2$ in the tight-binding limit.

Following the same procedure in Sec. II, one can alternatively
represent the Hamiltonian (\ref{Eq4:H1Dhaldane}) on a
two-dimensional lattice. Mathematically, by relabelling the mode
indices with respect to the honeycomb lattice, we can rewrite Eq.
(\ref{Eq4:H1Dhaldane}) as:
\begin{equation*}
H'_h = \sum_{\vec r} \left( \kappa c^\dagger_{B,\vec r} c_{A,\vec
r + \vec e_1} + \kappa c^\dagger_{B,\vec r} c_{A,\vec r + \vec
e_2} + \kappa c^\dagger_{B,\vec r} c_{A,\vec r + \vec e_3} + h.c.
\right)
\end{equation*}
\begin{equation*}
+  \sum_{\vec r} \left( \kappa' e^{-i\phi} c^\dagger_{A,\vec r}
c_{A,\vec r + \vec e_1 -\vec e_2} + \kappa' e^{i\phi}
c^\dagger_{B,\vec r} c_{B,\vec r + \vec e_1 -\vec e_2} + h.c.
\right)
\end{equation*}
\begin{equation}
+ \sum_i \left( \kappa c^\dagger_{B,(i+1,1)} c_{A,(i,N)} + \kappa'
e^{-i\phi} c^\dagger_{A,(i,N)} c_{A,(i+1,1)} + \kappa' e^{i\phi}
c^\dagger_{B,(i,N)} c_{B,(i+1,1)} + h.c. \right),
\label{Eq4:Hhaldane2}
\end{equation}
A specific example corresponding to 12 modes per ring, and with $N
= 4$ in Eq. (\ref{Eq4:Tbetween}), is shown in Figure
\ref{Fig:haldane}(e). This Hamiltonian therefore represents a
finite strip of a system described by the Haldane Hamiltonian,
placed on the surface of a cylinder with a twisted boundary
condition (Figure \ref{Fig:haldane}(f)). The system has two edges
at the end of the cylinder. Therefore, we expect that one-way edge
modes can exist on these edges.

\section{Realistic realization in ring resonators under dynamic modulations}

In the previous section we have provided a discussion for the
construction of a Haldane model in synthetic space. In this
section, we provide a detailed discussion of a possible
experimental implementation. In particular, in the previous
section we have described the effect of modulators in a simple
tight-binding model, which is valid only when the modulation
strength is sufficient small. In this section, we provide a more
realistic description of the field propagation inside the ring
under dynamic modulation. We also provide experimental signatures
of one-way edge states in the synthetic space. We show that such
one-way edge states can be probed by coupling input/output
waveguides to the system, and by analyzing the resulting output
spectrum.

\begin{figure}[h]
\centering
\includegraphics[width=1\linewidth]{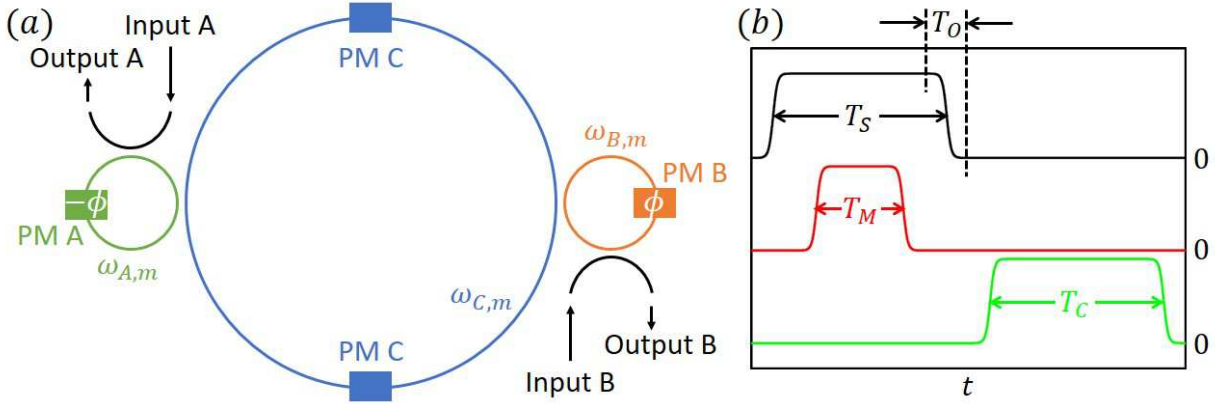}
\caption{(a) Same as Figure \ref{Fig:haldane}(c), except with
external waveguides coupling to rings $A$ and $B$. (b) The time
sequence for the input source (in black curve) with the temporal
width $T_S$, the modulation (in red curve) with $T_M$, and the
signal collection (in green curve) with $T_C$ in the proposed
experiment. All of these characteristic time windows have a
turn-on/off time $T_O$.\label{Fig:rings}}
\end{figure}

To simulate the evolution of the resonant modes in both the rings
$A$ and $B$, we can expand the electric field inside the ring as
\cite{hausbook}
\begin{equation}
E_{A(B)} (t,r_\bot,z)= \sum_m
\mathcal{E}_{A(B),m}(t,z)E_{A(B),m}(r_\bot)e^{i\omega_{A(B),m} t}
, \label{Eq:modeexpand}
\end{equation}
where the propagation direction inside the ring is given by $z$,
$r_\bot$ denotes the direction perpendicular to $z$,
$E_{A(B),m}(r_\bot)$ is the modal profile for the ring $A(B)$, and
$\mathcal{E}_{A(B),m}(t,z)$ gives the modal amplitude associated
with the $m$-th resonant mode. With this expansion, the Maxwell's
equations can then be converted into a set of dynamic equations
for the modal amplitudes $\mathcal{E}$. The effect of the
modulators is taken into account as a discontinuity condition for
$\mathcal{E}$ at the location of modulators in the ring. The
coupling between the rings are implemented using the coupling
matrix (\ref{Eq7:Tcouple}) at the location of the waveguide
couplers. Rings $A$ and $B$ couple to external waveguides $A$ and
$B$ respectively with the coupling strength $\gamma_w$ with the
same form of coupling matrix in Eq. (\ref{Eq7:Tcouple}). This
procedure is described in details in Refs.
\cite{yuanOL,yuanOptica} and we briefly summarize it in the
Appendix.

In the simulation, we assume that both rings $A$ and $B$ support
66 resonant modes ($m=1,2,\ldots,66$). The group velocity
dispersion beyond this range provides the ``boundary'' along the
frequency axis \cite{yuanOL}. In Eq. (\ref{Eq4:Tbetween}), which
describes the modulators inside the ring $C$, we choose $N=11$.
Therefore, our design gives a two-dimensional $6 \times 11$
lattice in the synthetic space. The parameters for the system in
Figure \ref{Fig:rings}(a) are set as $\kappa_1 = 0.2$,
$\kappa'=0.001$, $\gamma = 0.1$, and $\gamma_w = 0.007$. The
source is injected into the system and the signal is detected
through the external waveguides.

In the proposed set of experiments, we turn on the source first.
After the source is turned on, we turn on the modulation over a
duration of $T_M$, and then turn the modulation off. Afterwards we
turn off the source. The duration of the source is $T_S$. After
the source is turned off, we collect the output signal over a
duration of $T_C$ in both waveguides $A$ and $B$. We then Fourier
transform the collected signal to obtain the output spectrum. We
smooth all the process of turning on and off over a period of
$T_O$. A schematic showing this sequence of events is shown in
Figure \ref{Fig:rings}(b). In the following simulations, we choose
$T_C = 800$ $n_g L/c$ and $T_O=200$ $n_g L/c$. We choose $T_M = T,
2T, 3T, 4T$, respectively, where $T=4000$ $n_g L/c$. And we set
$T_S = T_M + 2T_O$ for each $T_M$. We excite the system by an
input source at the single frequency $\omega_{A,m=6}$ through the
waveguide $A$. We collect the output $E^{O}_{A(B)} (t)$ at the
external waveguides $A$ and $B$. The output spectra are then
calculated by Fourier transform $E^{O}_{A(B)} (\omega) \equiv
\int_{T_C} dt E^{O}_{A(B)} (t) e^{-i\omega t}$. The output spectra
typically consist of peaks at the frequencies $\omega_{A),m}$ and
$\omega_{B,m}$ for the two output waveguides, respectively.

\begin{figure}[h]
\centering
\includegraphics[width=\linewidth]{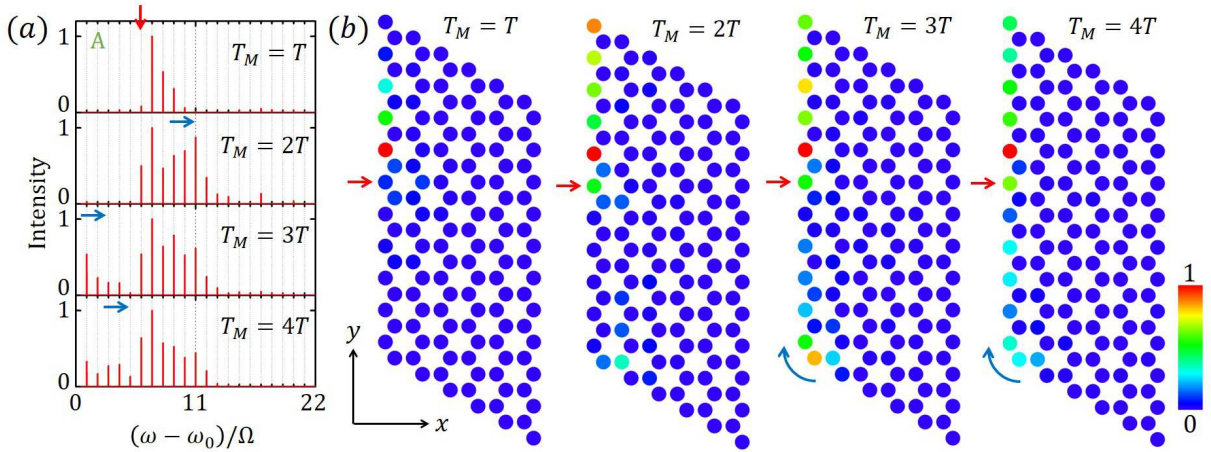}
\caption{(a) Output spectra in external waveguide $A$ for the
system as shown in Figure \ref{Fig:rings}, with $T_M =
T,2T,3T,4T$, respectively. The spectra are normalized to the range
$[0,1]$. Only 22 resonant modes are plotted. The amplitudes for
other modes are close to zero. (b) Output spectra are re-plotted
in the synthetic two-dimensional space for $T_M = T$, $2T$, $3T$,
and $4T$, respectively. Red arrows show the frequency of the input
excitation. Blue arrows denote the unidirectional transport in the
synthetic space.  \label{Fig:ringsim}}
\end{figure}

The simulation results are shown in Figure \ref{Fig:ringsim}. As
we increase $T_M$, the output spectra show the signature of
one-way frequency transport in the synthetic space. At $T_M = T$,
only frequency components higher than the input frequency at the
6th resonant mode are generated. At $T_M = 2T$, the 11th resonant
mode is significantly excited. At $T_M = 3 T$, the first resonant
mode is excited. The spectrum then again shifts to the higher
frequency direction at $T_M = 4 T$. This process can be also
represented on the synthetic lattice  as shown in Figure
\ref{Fig:ringsim}(b), where we plot the different frequency
components of the $A$ and $B$ output waveguides on a honeycomb
lattice as described in Section III. In this synthetic lattice, at
$T_M = T$, the light propagates along the left edge
unidirectionally upward. At $T_M = 2T$, it reaches the upper end
of the lower edge. Due to the twisted connections between the
upper and lower boundaries, once the light reaches to the
left-upper corner, it transports back to the lower edge, as seen
at $T_M = 3T$. After that, it again travels upward, as seen in
$T_M = 4T$. This trajectory corresponds to one-way transport of
light along the left boundary in a finite structure with a twisted
boundary condition (Figure \ref{Fig:haldane}(f)), and represent
the signature that a Haldane model can be realized in synthetic
dimensional using three resonators.

To accomplish the proposed experiment above, one can use either
silicon or lithium niobate ring resonators with radius of
approximately $100$ $\mathrm{\mu m}$ with a free-spectral-range
$\Omega \sim 100$ GHz. For modulation, the value of $\kappa_1$ and
$\kappa'$ corresponds to the efficiency for conversion to other
frequencies after an incident wave with a certain frequency passes
through the modulator. Our choice of $\kappa_1$ and $\kappa'$
above corresponds to a conversion efficiency of $2\%$ and
$5\times10^{-5}\%$, which is reasonable for electro-optic
modulators \cite{tzuangnp14,tzuang14,wang17,reed14}. To observe
the signal with the modulation in a time scale at the order of
$4T$ in Figure \ref{Fig:ringsim} requires a loss $<10^{-4}$ in
energy per round trip. This corresponds to a ring resonator of a
quality factor up to $~10^7 - 10^8$, which is challenging but
possibly achievable under the current photonic technology
\cite{ilchenko04,savchenkov08,wang15}. Following this proposal,
the entire structure fits into a $\sim 1 \times 1$ mm$^2$ photonic
chip.


\section{Conclusion}

In summary, we have proposed that one can achieve arbitrary
dimensions in a single dynamically modulated ring resonator by
choosing a particular modulation format. In addition, we find that
one can construct the Haldane model in a synthetic frequency space
with three dynamically modulated rings. Our study suggests that a
``zero-dimensional'' geometric structure can exhibit topological
physics in two or even higher dimensions. The dynamic of light in
this synthetic space may also provide new capabilities for the
control of the spectrum of light.

\begin{acknowledgments}
This work is supported by U.S. Air Force Office of Scientific
Research Grants No. FA9550-12-1-0488, and  FA9550-17-1-0002, as
well as the U. S. National Science Foundation Grant No.
CBET-1641069.
\end{acknowledgments}

\newpage

\appendix

\section*{Appendix --- simulation method based on the modal
expansion of the electric field}

\renewcommand{\theequation}{A-\arabic{equation}}
\setcounter{equation}{0}

We study the evolution of the resonant modes in both the rings $A$
and $B$. The modal amplitude $\mathcal{E}_{A(B),m}(t,z)$ at the
$m$-th resonant mode defined in Eq. (\ref{Eq:modeexpand}) obeys
the equation \cite{hausbook2,yuanOL2,yuanOptica2}
\begin{equation}
\left(\frac{\partial}{\partial z} + i\beta (\omega_{A(B),m})
\right) \mathcal{E}_{A(B),m}(t,z) - \frac{n_g(\omega_{A(B),m})}{c}
\frac{\partial}{\partial t}\mathcal{E}_{A(B),m}(t,z) = 0
\label{Eq:evolution}
\end{equation}
under the slowly-varying-envelope approximation and the boundary
condition $\mathcal{E}_{A(B),m}(t,z+L) =
\mathcal{E}_{A(B),m}(t,z)$. Here $n_g$ is the group index and is
assumed to be constant in the regime with the zero group velocity
dispersion.

The resonant modes in $A$ and $B$ are off resonance from the modes
in $C$, since light is injected into ring $C$ only through rings
$A$ and $B$, we therefore expand the field inside ring $C$ with
only the field components that are resonant in the ring $A(B)$:
\begin{equation}
E_{C} (t,r_\bot,z)= \sum_m
\mathcal{E}_{CA,m}(t,z)E_{CA,m}(r_\bot)e^{i\omega_{A,m} t} +
\mathcal{E}_{CB,m}(t,z)E_{CB,m}(r_\bot)e^{i\omega_{B,m} t} .
\label{Eq:modeexpand2}
\end{equation}
The associated mode amplitude satisfies
\begin{equation}
\left(\frac{\partial}{\partial z} + i\beta (\omega_{A(B),m})
\right) \mathcal{E}_{CA(CB),m}(t,z) -
\frac{n_g(\omega_{A(B),m})}{c} \frac{\partial}{\partial
t}\mathcal{E}_{CA(CB),m}(t,z) = 0 \label{Eq:evolution2}
\end{equation}
under the slowly-varying-envelope approximation and
$\mathcal{E}_{CA(CB),m}(t,z+4L) = \mathcal{E}_{CA(CB),m}(t,z)$.

The modulators in the rings have transmission coefficients
described by Eqs. (\ref{Eq4:TAB}) and (\ref{Eq4:Tbetween}). At the
location of these modulators, the modal amplitudes satisfy
\cite{Saleh}:
\begin{equation*}
\mathcal{E}_{A(B),m} (t^+,z_{A(B)}) = J_0 (\kappa')
\mathcal{E}_{A(B),m} (t^-,z_{A(B)})
\end{equation*}
\begin{equation}
+ \sum_{q>0} J_q (\kappa') \left[ \mathcal{E}_{A(B),m-q}
(t^-,z_{A(B)}) e^{\pm i q\phi} + (-1)^q \mathcal{E}_{A(B),m+q}
(t^-,z_{A(B)}) e^{\mp iq\phi} \right], \label{Eq:phasemodAB}
\end{equation}
\begin{equation*}
\mathcal{E}_{CA,m} (t^+,z_{C}) = J_0^3 (\kappa_1)
\mathcal{E}^3_{CA,m} (t^-,z_{C})
\end{equation*}
\begin{equation}
 + J_0^2 (\kappa_1) J_1 (\kappa_1) \mathcal{E}^2_{CA,m} (t^-,z_{C}) \left[ \mathcal{E}_{CB,m} (t^-,z_{C}) -
\mathcal{E}_{CB,m+1} (t^-,z_{C}) +\mathcal{E}_{CB,m-N} (t^-,z_{C})
\right], \label{Eq:phasemodCA}
\end{equation}
\begin{equation*}
\mathcal{E}_{CB,m} (t^+,z_{C}) = J_0^3 (\kappa_1)
\mathcal{E}^3_{CB,m} (t^-,z_{C})
\end{equation*}
\begin{equation}
- J_0^2 (\kappa_1) J_1 (\kappa_1) \mathcal{E}^2_{CB,m} (t^-,z_{C})
\left[ \mathcal{E}_{CA,m} (t^-,z_{C}) - \mathcal{E}_{CA,m-1}
(t^-,z_{C}) + \mathcal{E}_{CA,m+N} (t^-,z_{C}) \right],
\label{Eq:phasemodCB}
\end{equation}
where the rotating wave approximation is used. $J_q$ is the $q$-th
order Bessel function and $t^\pm = t+0^\pm$. $z_{A,B,C}$ denote
the position of the modulators in each ring in Figure
\ref{Fig:rings}(a). In Eq. (\ref{Eq:phasemodAB}), which describes
the modulators in rings $A$ and $B$, we expand the transmission
coefficient as described by Eq. (\ref{Eq4:TAB}) to all orders of
its Fourier coefficients. On the other hand, in Eqs.
(\ref{Eq:phasemodCA}) and (\ref{Eq:phasemodCB}), which describes
the modulators in ring $C$, we expand the transmission coefficient
of Eq. (\ref{Eq4:Tbetween}) and keep only the lowest frequency
components to the 1st order. We also keep only the contributions
from the field components that are resonant in the ring $A(B)$.
This expansion is reasonable since ring $C$ is non-resonant.

The coupling between rings $A$ or $B$ and ring $C$ is described
by:
\begin{equation}
\mathcal{E}_{A(B),m} (t^+,z'_{A(B)}) = \sqrt{1-\gamma^2}
\mathcal{E}_{A(B),m} (t^-,z'_{A(B)}) - i\gamma
\mathcal{E}_{CA(CB),m} (t^-,z'_{CA(CB)}), \label{Eq:ringcouple1}
\end{equation}
\begin{equation}
\mathcal{E}_{CA(CB),m} (t^+,z'_{CA(CB)}) = \sqrt{1-\gamma^2}
\mathcal{E}_{CA(CB),m} (t^-,z'_{CA(CB)}) - i\gamma
\mathcal{E}_{A(B),m} (t^-,z'_{A(B)}), \label{Eq:ringcouple2}
\end{equation}
where $\gamma$ is the coupling strength between rings. The $z'$ s
denote the positions of the coupler in the rings as specfied by
the subscript.  The coupling between the rings $A$($B$) and the
external waveguides $A$($B$) is given by
\begin{equation}
\mathcal{E}_{A(B),m} (t^+,z''_{A(B)}) = \sqrt{1-\gamma_w^2}
\mathcal{E}_{A(B),m} (t^-,z''_{A(B)}) - i\gamma_w E^{I}_{A(B),m}
(t^-,z''_{A(B)}), \label{Eq:ringwaveguide1}
\end{equation}
\begin{equation}
E^{O}_{A(B),m} (t^+,z''_{A(B)}) = \sqrt{1-\gamma_w^2}
E^{I}_{A(B),m} (t^-,z''_{A(B)}) - i\gamma_w \mathcal{E}_{A(B),m}
(t^-,z''_{A(B)}), \label{Eq:ringwaveguide2}
\end{equation}
where $E^{I/O}_{A(B),m}$ is the input/output amplitude of the
field component at the frequency mode $\omega_{A(B),m}$ in the
external waveguides $A$($B$) and $\gamma_w$ is the coupling
strength. The $z''$ s denote the position of the coupler between
the external waveguides and the rings.

\end{document}